\documentstyle[aps,multicol,epsf]{revtex}
\begin{document}
\preprint{}
\draft
\title{Effect of Anisotropy on the Localization in a Bifractal System}
\author{P. H. Song and Doochul Kim}
\address{Department of Physics and Center for Theoretical Physics, Seoul 
National University, Seoul 151--742, Korea}
%\date{\today}
%\ \hspace{5in}{\normalsize SNUTP }\\
\maketitle
\begin{abstract}
Bifractal is a highly anisotropic structure where planar fractals are 
stacked to form a 3-dimensional lattice.  The localization lengths along 
fractal structure for the Anderson model defined on a bifractal are 
calculated.  The critical disorder 
and the critical exponent of the localization lengths are obtained from
the finite size scaling behavior.  The numerical results are in a good
agreement with previous results which have been obtained from the 
localization lengths along the perpendicular direction. 
This suggests that the anisotropy of the embedding lattice structure is
irrelevant to the critical properties of the localization.
\end{abstract}


\begin{multicols}{2}
\narrowtext

Recently, there has been much attention \cite{bha,qim,roj,abri,xue,zam} to 
the critical properties of 
the localization in anisotropic system.  For a 3-dimensional cubic system
with anisotropic hopping matrix, it seems now that there is a general 
agreement \cite{zam} that both the critical disorder ($W_c$) and the 
critical exponent ($\nu$) of the localization length are independent of 
direction of measurement and that such model belongs to the universality 
class of the isotropic Anderson model.  

On the other hand, bifractals \cite{shi,sch} are constructed by stacking 
planar fractal lattices along the $z$-direction so that they are of 
Euclidean structure only in the $z$-direction.
Anisotropy in these systems arises from the lattice structure 
itself.  Therefore, it is by no means obvious whether the critical properties 
obtained from the localization lengths along the $xy$-plane 
($\lambda_{xy}$) are the same as those from the localization lengths 
along the $z$-axis ($\lambda_z$).  There is a possibility that even a 
mobility edge, one of useful concepts in the localization theory, may not 
exist in this intrinsically anisotropic system.  

Therefore, in this Report, we study the critical disorder and the 
critical exponent along fractal structure, i.e. along the $xy$-plane, 
for a bifractal system.  Our results are in excellent agreement with those 
obtained from $\lambda_z$'s, which have been reported in previous studies 
\cite{sch,gru}, suggesting that the anisotropy of the embedding lattice 
structure is irrelevant to the critical properties of the localization.  In our 
study, one of the bifractals introduced in the Ref. [8] is used as 
the model.  We consider the Anderson Hamiltonian given as 
\begin{equation}
{\cal H} = \sum_i \epsilon_i|i\rangle\langle i| +
\sum_{\langle i,j\rangle} V(|i\rangle\langle j|+|j\rangle
\langle i|),
\end{equation}
where the random site energies $\epsilon_i$ are chosen from a 
box distribution of width $W$.  The hopping energies $V$ are set to 1 
throughout this work and $\langle i,j\rangle$ denotes the sum over nearest 
neighbor pairs of sites on the bifractal lattice.  In Fig.~1(a), the 
lattice is schematically depicted.  
The cross section perpendicular to the $z$-axis is a variant of 
the Sierpinski gasket and $L$ is the number of the fractal lattices that 
have been stacked.  The number of the iteration processes for the fractal 
lattice is denoted by $n$, e.g. $n$ = 2 for Fig.~1(a).  This is 
exactly the model called as the Bifractal I in the Ref. 
[8].  The Green's function coupling two corner sites of the largest triangle, 
{\bf r} and ${\bf r^{\prime}}$, is denoted as $G_{{\bf r},{\bf r'}}(n,L)$.  
Then the localization length along the $xy$-plane, $\lambda_{xy}(L)$, can be 
defined as follows;
\begin{equation}
\frac{1}{\lambda_{xy}(L)} = -\lim_{n\rightarrow \infty} 
\frac{1}{|{\bf r} - {\bf r'}|} \mbox{log}|G_{{\bf r},{\bf r'}}(n,L)|,
\end{equation}
where {\bf r} and ${\bf r^{\prime}}$ have the same $z$-coordinates.

The main point of the calculation is to find $G_{{\bf r},{\bf
r'}}(n,L)$'s for sufficiently large value of $n$, which means that one
should calculate elements of the inverse of a very large
random matrix, $(E - {\cal H})$.  This is essentially the same 
problem as encountered in the transfer matrix method for 
quasi-1-dimensional systems \cite{mac}.  However a different recursive 
algorithm should be devised since we are considering a ``quasi-2-dimensional 
system".

One can handle the problem by decimating recursively the amplitudes of 
the sites characterized by the largest iteration number, as shown in 
Fig.~1(b).  Following scheme is for the case of 
$L=1$ but the extension of the method for $L \geq 2$ is straightforward.
Let {\bf x} ({\bf y}) be a vector the elements of which are the 
amplitudes of the sites represented by the solid (empty) circles in 
Fig.~1(b).  Then a matrix equation for {\bf x} and {\bf y} 
can be constructed in the form,
\begin{equation}
\left( \begin{array}{cc}
H_x(E) & V_{xy}(E)\\
V_{xy}^t(E) & H_y(E)\\
\end{array} \right)
\left( \begin{array}{c}
{\bf x}\\
{\bf y}\\
\end{array} \right)
= 
\left( \begin{array}{c}
{\bf z}\\
{\bf 0}\\
\end{array} \right)
\end{equation}
where $H_x(E), H_y(E)$ and $V_{xy}(E)$ are matrices and contributions from 
remaining sites other than shown in Fig.~1(b) are contained in a vector 
{\bf z}.  The vector {\bf 0} in the right hand side of Eq.~(3) represents the 
fact that {\bf y} is directly coupled only with {\bf x}.  Initially, when 
the sites have been indexed as in Fig.~1(b), the explicit forms of the three 
matrices are as follows; (i) $H_{x,ij} = \delta_{ij} (E-\epsilon_{x_i})$,
(ii) $H_{y,ii} = E-\epsilon_{y_i}, H_{y,12} = H_{y,21} = H_{y,13}
= H_{y,31} = H_{y,25} = H_{y,52} = H_{y,34} = H_{y,43} = H_{y,46} = H_{y,64} 
= H_{y,56} = H_{y,65} = -V$ and $H_{y,ij} = 0$ otherwise, (iii) $V_{xy,11} = 
V_{xy,12} = V_{xy,23} = V_{xy,24} = V_{xy,35} = V_{xy,36} = -V$ and $V_{xy,ij}
= 0$ otherwise.  Eliminating the amplitudes of the internal sites, 
i.e. {\bf y}, we get 
\begin{equation}
(H_x-V_{xy}H_y^{-1}V_{xy}^t)\ {\bf x} = {\bf z}.
\end{equation}
By performing the decimation process of Eq.~(4) for every triangle consisting
of nine sites, the number of the whole eigenvalue equations reduces by a factor
of three.  The $3 \times 3$ matrix in the left hand side of Eq.~(4) defines the 
renormalized hopping energies within the smallest triangles of the new lattice 
and the renormalized on-site terms.  Since the hopping energies are modified only
within the smallest triangle one can cast the matrix equation for the remaining 
sites again in the form of Eq.~(3).  It should be also noted that at the first
step of the iteration, the elements of $V_{xy}$ are independent of $E$ as
can be seen from iii), while after the decimation process, i.e. Eq.~(4),
they become functions of $E$, in general.  Therefore, the problem has been reduced 
to another eigenvalue equation problem on the Sierpinski gasket with the 
iteration number smaller than the original by 1.  Now we can iterate the 
above procedure until 3 linear equations for the amplitudes of the 3 outermost 
sites are left in case of $L = 1$.  For general $L$, we have $3 \times L$ 
linear equations instead of 3.  Then the inverse of the matrix constructed 
by the coefficients of the amplitudes is the Green's function for the outermost 
sites.   

In principle, this algorithm can be iterated infinitely.  However,
one of the technical problems in real calculation is that some off-diagonal
elements of the matrix becomes very small compared with, say, ${\cal O}(1)$
diagonal elements, as the iteration proceeds.  In general, these small 
elements contain the essential information for our problem since
they are directly related to the Green's function connecting two different 
sites far away from each other.  For example, let's assume that the 
localization length
is of order unity and the distance between two sites is, say, 1000 $\times$
(lattice constant).  Then the Green's function connecting the two sites is 
of order $e^{-1000} \sim 10^{-430}$ while the diagonal elements of the
Green's function are of order unity.  Therefore one is investigating an
asymptotic behavior of vanishingly small matrix elements while the order of
magnitudes of some other elements of the matrix is far much larger than them.
Direct manipulation of such matrix leads to loss of information on the smaller
elements. 
One of possible techniques to overcome this difficulty is to decompose the 
matrix into two parts as
\begin{equation}
A = A_0 + e^{-\alpha} A^{\prime},
\end{equation}
where the matrix elements of both $A_0$ and $A^{\prime}$ are within the 
range safely handled by computers.  $\alpha$ is a number and should be 
modified whenever the matrix is manipulated.  When a matrix is decomposed 
as the above, the inverse of the matrix can be calculated by the formula 
\begin{equation}
(A_0 + e^{-\alpha} A^{\prime})^{-1} = A_0^{-1} +  A_0^{-1} \sum_{n=1}^\infty 
(-e^{-\alpha} A^{\prime} A_0^{-1})^n.
\end{equation}
For our calculation, it turns out to be sufficient to retain terms up to 
second order in $e^{-\alpha}$.

We calculate $\lambda_{xy}(L)$'s for $E = 0$ and several 
values of $W$ in the range $4.0 \leq W \leq 9.0$.  
In our calculation, the Sierpinski gasket with $n = 11$ ($3^{12}$ 
sites/cross section) is used and $L$ is varied within the range 
$3 \leq L \leq 15$.  For a given set of parameters, configurational 
averages are performed over 4 $\sim$ 70 different realizations to control 
the uncertainty of $\lambda_{xy}(L)$ within 1 $\%$.  Periodic boundary 
condition is imposed in the $z$-direction.  The results are 
shown in Fig.~2.  As $L$ increases, the renormalized localization 
length, $\lambda_{xy}(L)/L$, increases for smaller values of $W$, 
while for larger values it keeps decreasing.  This implies that in the 
macroscopic limit, i.e., $L \rightarrow \infty$, there exists a 
transition from an extended state to a localized state as $W$ varies, 
e.g., from 5.0 to 6.0.  Estimates of $W_c$
and $\nu$ are obtained by fitting the data to the scaling form
\begin{equation}
\log(\lambda_{xy}(L)/L) = a + b (W-W_c) L^{1/\nu}.
\end{equation}
where $a$ and $b$ are constants.
Several sets of data for $4.5 \leq W \leq 7$ and $7 \leq L \leq 15$, 
have been fitted to the equation (7) and finally we get $W_c = 5.79 \pm 
0.04$ and $\nu = 2.92 \pm 0.14$.  The errors are the 
dispersions between different sets of data.  The scaling plot with these
values is shown in Fig.~3.  Our numerical results are in good agreement 
with those obtained from the analysis of $\lambda_z$'s, {\it i.e.} 
$W_c = 5.8$ and $\nu = 3.0 \pm 0.2$\cite{gru} for the same model.  This 
supports the idea that $W_c$ and $\nu$ are independent of the direction 
of measurement for this bifractal system.

That $W_c$'s are the same along the two directions indicates that there 
exists a well-defined mobility edge in spite of the intrinsic anisotropy in 
this model.  In addition we can expect that the 
anisotropy is irrelevant to the localization transition in a somewhat 
strong sense, that is, {\it the critical properties of the localization 
is insensitive not only to the anisotropy in the energy parameters but 
also to that of the embedding lattice structure}.  

We are  thankful to M. Schreiber and H. Grussbach for sending their
numerical data.  This work has 
been supported by the Korea Science and Engineering Foundation through 
the Center for Theoretical Physics and by the Ministry of Education through 
BSRI both at Seoul National University.

\begin{figure}
\centerline{\epsfxsize=7cm \epsfbox{fig1.epsi}}
\vspace{4mm}
\caption{
(a) Schematic diagram of the lattice with the iteration 
number $n$ = 2 and $L$ = 5.  Each vertex is a lattice site of the 
Hamiltonian given by 
Eq.~(1).  (b) Schematic diagram of the decimation process.  The amplitudes of the
sites represented by the empty circles are eliminated to modify the forms of 
the equations for the sites represented by the solid circles.
}
\end{figure}
\noindent
\begin{figure}
\centerline{\epsfxsize=9cm \epsfbox{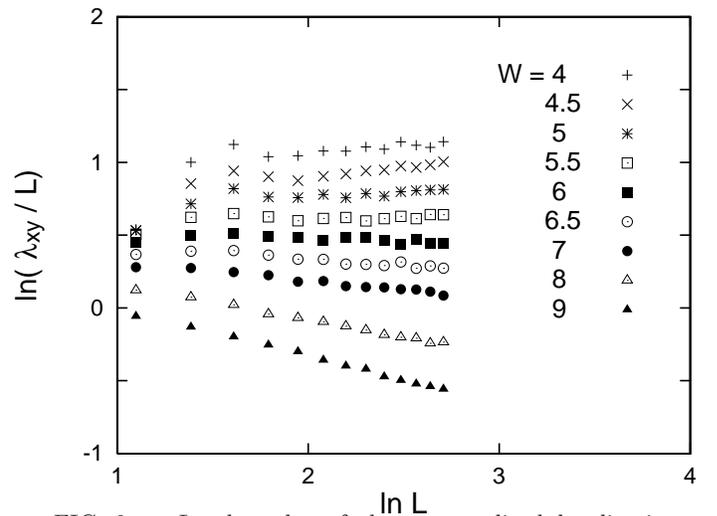}}
\caption{
Log-log plot of the renormalized localization length 
along the Sierpinski gasket as a function of $L$ for various values 
of $W$.  The uncertainty of each data point is less than the symbol 
size.
}
\end{figure}
\noindent
\begin{figure}
\centerline{\epsfxsize=9cm \epsfbox{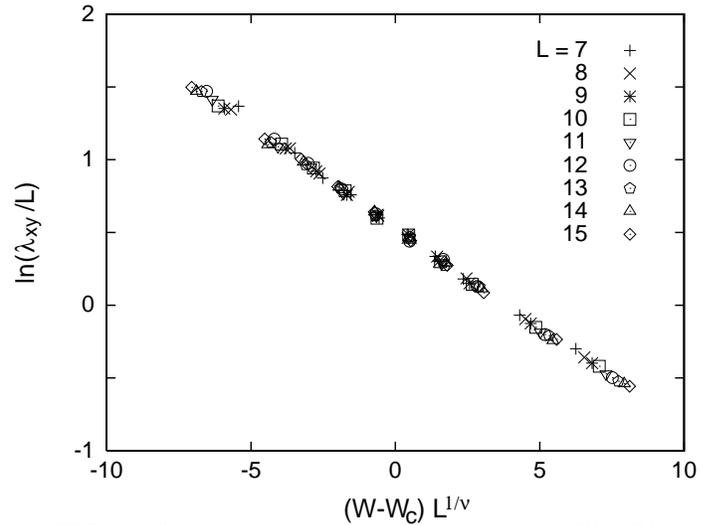}}
\caption{ 
Scaling behavior of $\log(\lambda_{xy}/L)$ versus 
$(W-W_c)\ $ $ L^{1/\nu}$ with $W_c$ = 5.79 and $\nu$ = 2.92.
}
\end{figure}
\noindent
\end{multicols}
\end{document}